\documentclass{article}
\usepackage{proceedings}
\usepackage{graphicx}
\usepackage{latexsym}
\newtheorem{theorem}{Theorem}
\newtheorem{definition}{Definition}
\newtheorem{example}{Example}
\newtheorem{proposition}{Proposition}

\newcommand{\implies}{\Rightarrow}
\newcommand{\proof}{Proof.}
\begin{document}

\title{Wat is a Joint Goal?\\Games with Beliefs and Defeasible Desires}
\author{{\bf Mehdi Dastani} \\ Institute of Information and Computing Sciences\\
Utrecht University \\ email: mehdi@cs.uu.nl \\ \And  {\bf Leendert
van der Torre} \\ Department of Artificial Intelligence \\ Vrije
Universiteit Amsterdam \\ email: torre@cs.vu.nl }

\maketitle

\begin{abstract}
In this paper we introduce a qualitative decision and game theory
based on belief (B) and desire (D) rules. We show that a group of
agents acts as if it is maximizing achieved joint goals.
\end{abstract}

\section{Introduction}

One of the main problems in agent theory is the distinction
between formal theories and tools developed for individual
autonomous agents, and theories and tools developed for multi
agent systems. In the social sciences, this distinction is called
the micro-macro dichotomy. The prototypical example is the
distinction between classical decision theory based on the
expected utility paradigm (usually identified with the work of
Neumann and Morgenstern \cite{neumann} and Savage \cite{savage})
and classical game theory (such as the work of Nash and more
recently the work of Axelrod). Whereas classical decision theory
is a kind of optimization problem (maximizing the agent's expected
utility), classical game theory is a kind of equilibria analysis.

There are several approaches in practical reasoning (within
philosophy), cognitive science and artificial intelligence to
bring the micro and macro description together. The basic idea is
two-fold:
\begin{enumerate}
\item
The decision making of individual autonomous agents is described
in terms of other concepts than maximizing utility. For example,
since the early 40s there is a distinction between classical
decision theory and artificial intelligence based on utility
aspiration levels and goal based planning (as pioneered by Simon
\cite{simon}). Cognitive theories are typically based on vague
concepts from folk psychology like beliefs, desires and
intentions.
\item
The decision making of a group or society of agents is described
in terms of concepts generalized from those used for individual
agents, such as joint goals, joint intentions, joint commitments,
etc. Moreover, also new concepts are introduced at this social
level, such as norms (a central concept in most social theories).
\end{enumerate}
It is still an open problem how the micro-macro dichotomy of
classical decision and game theory is related to the micro-macro
dichotomy of these alternative theories. Has or can the micro and
macro level be brought together by replacing classical theories by
alternative theories? Before this question can be answered, the
relation between the classical and alternative theories has to be
clarified. Doyle and Thomason \cite{doyle:aimagazine99} argue that
classical decision theory should be reunited with alternative
decision theories in so-called qualitative decision theory (QDT),
which studies qualitative versions of classical decision theory,
hybrid combinations of quantitative and qualitative approaches to
decision making, and decision making in the context of artificial
intelligence applications such as planning, learning and
collaboration. Qualitative decision theories have been developed
based on beliefs (probabilities) and desires (utilities) using
formal tools such as modal logic \cite{boutilier:kr94} and on
utility functions and knowledge \cite{lang:ecai96}. More recently
these beliefs-desires models have been extended with intentions or
BDI models \cite{cohen,rao:kr91}.

In this paper we introduce a rule based qualitative decision and
game theory, based on belief (B) and desire (D) rules. We call an
individual autonomous agent which minimizes its unreached desires
a BD rational agent. We define goals as a set of formulas which
can be derived by beliefs and desires in a certain way, such that
BD rational agents act as if they maximize the set of achieved
goals, and agents maximizing their sets of achieved goals are BD
rational. Moreover, groups of agents which end up in equilibria
act as if they maximize joint goals.

Like classical decision theory but in contrast to several
proposals in the BDI approach \cite{cohen,rao:kr91}, the theory
does {\em not} incorporate decision processes, temporal reasoning,
and scheduling.

The layout of this paper is as follows. Section \ref{sec:qdt}
develops a qualitative logic of decision. This logic tells us what
the optimal decision is, but it does not tell us how to find this
optimal decision. Section \ref{sec:gdt} considers the AI solution
to this problem \cite{simon,newell}: break down the decision
problem into goal generation and goal based decisions.

\subsection{QDT and NMR}

Qualitative decision theory is related to non-monotonic logic
(``qualitative'') and to reasoning about uncertainty (``decision
theory'' formalizes decision making under uncertainty). However,
they are based on different disciplines. According to a
distinction made by Aristotle, non-monotonic reasoning and
reasoning about uncertainty formalize theoretical (or conclusion
oriented) reasoning, whereas qualitative decision theory
formalizes practical (or action oriented) reasoning.

Thomason \cite{thomason:ijcai99notes} observes that classical
decision theory neglects the issue of `correct inference', and
that the absence of a logical theory of practical reasoning is
largely due to the unavailability of appropriate inference
mechanisms. To handle even the simplest cases of practical
reasoning, it is essential to have a reasoning mechanism that
allows for practical conclusions that are non-monotonic in the
agent's beliefs.

Where classical decision theory is based on probabilities and
utilities, qualitative decision theory is based on beliefs and
desires. In a modal approach, where the beliefs and desires are
represented by modalities $B$ and $D$ respectively, and an action
operator by the modal operator $Do$, we may have for example that:
$$B(thirsty), D(drink) \vdash Do(go-to-pub)$$
$$B(thirsty), D(drink), B(pub-closed) \vdash Do(go-to-shop)$$

A drawback of such a modal logic approach is that on the one hand
modal logic is notorious for its problems with formalizing
conditionals or rules, and practical reasoning on the other hand
is usually seen as a kind of rule based reasoning. In this paper
we therefore do not use modal logic but we use a rule based
formalism.

\section{A qualitative decision and game theory}
\label{sec:qdt}

The qualitative decision and game theory introduced in this
section is based on sets of belief and desire rules. We define an
agent system specification, we show how we can derive a game
specification from it, and we give some familiar notions from game
theory such as Pareto efficient decisions (choosing an optimal
decision) and Nash equilibria. First we consider the logic of
rules we adopt.

\subsection{Logic of rules}

The starting point of any theory of decision is a distinction
between choices made by the decision maker and choices imposed on
it by its environment. For example, a software upgrade agent
(decision maker) may have the choice to upgrade a computer system
at a particular time of the day. The software company
(environment) may in turn allow/disallow such a upgrade at a
particular time. Let $S=\{\alpha_1,\ldots,\alpha_n\}$ be the
society or set of agents, then we therefore assume $n$ disjoint
sets of propositional atoms: $A=A_1\cup \ldots\cup
A_n=\{a,b,c,\ldots\}$ (agents' decision variables
\cite{lang:ecai96} or controllable propositions
\cite{boutilier:kr94}) and $W=\{p,q,r,\ldots\}$ (the world
parameters or uncontrollable propositions). In the sequel we
consider each decision maker as entities consisting of defeasible
rules. Such a decision maker generates its decisions by applying
subsets of defeasible rules to its input. This results in the
so-called conditional mental attitudes \cite{broersen:aa01}.
Before we proceed some notations will be introduced.

\begin{itemize}
\item
$L_{A_i}$, $L_W$ and $L_{AW}$ for the propositional languages
built up from these atoms in the usual way, and variables $x$,
$y$, \ldots\ to stand for any sentences of these languages.
\item
$Cn_{A_i}$, $Cn_W$ and $Cn_{AW}$ for the consequence sets, and
$\models_{A_i}$, $\models_W$ and $\models_{AW}$ for
satisfiability, in any of these propositional logics.
\item
$x\implies y$ for an ordered pair of propositional sentences
called a rule.
\item
$E_R(T)$ for the $R$ extension of $T$, as defined in Definition
\ref{def:iol} below.
\end{itemize}

In our framework the generation of decisions are formalized based
on the notion of extension. In particular, the decision of an
agent, which is specified by a set of defeasible rules $R$ and has
the input $T$, is the extension calculated based on $R$ and $T$.
This is formalized in the following definition.

\begin{definition}[Extension] \label{def:iol} Let $R\subseteq L_{AW}\times
L_{AW}$ be a set of rules and $T\subseteq L_{AW}$ be a set of
sentences. The consequents of the $T$-applicable rules are:
$$R(T)=\{y\mid x\implies y \in R, x\in T\}$$
and the $R$ extension of $T$ is the set of the consequents of the
iteratively $T$-applicable rules:
$$E_R(T)=\cap_{T\subseteq X,R(Cn_{AW}(X))\subseteq X}X$$
\end{definition}

We give some properties of the $R$ extension of $T$ in Definition
\ref{def:iol}. First note that $E_R(T)$ is {\em not} closed under
logical consequence. The following proposition shows that $E_R(T)$
is the smallest superset of $T$ closed under the rules $R$
interpreted as inference rules.

\begin{proposition}
Let
\begin{itemize}
\item
$E_R^0(T)=T$
\item
$E_R^i(T)=E_R^{i-1}(T)\cup R(Cn_{AW}(E_R^{i-1}(T)))$ for $i>0$
\end{itemize}
We have $E_R(T)=\cup_0^{\infty}E_R^{i}(T)$.
\end{proposition}

Usually, an decision making agent is required to preserve its
decisions under the growth of inputs. The following proposition
shows that $E_R(T)$ is monotonic.

\begin{proposition}
We have $R(T)\subseteq R(T\cup T')$ and $E_R(T)\subseteq E_R(T\cup
T')$.
\end{proposition}

Monotonicity is illustrated by the following example.
\begin{example}
Let $R=\{\top\implies p,a\implies \neg p\}$ and $T=\{a\}$, where
$\top$ stands for any tautology like $p\vee\neg p$. We have
$E_R(\emptyset)=\{p\}$ and $E_R(T)=\{a,p,\neg p\}$, i.e. the $R$
extension of $T$ is inconsistent.
\end{example}

Of course, allowing inconsistent decisions may not be intuitive.
We are here concerned about possible decisions rather than
reasonable or feasible decisions. Later we will define reasonable
or feasible decisions by excluding inconsistent decisions.

\subsection{Agent system specification}

An agent system specification given in Definition \ref{def:ds}
contains a set of agents and for each agent a description of its
decision problem. The agent's decision problem is defined in terms
of its beliefs and desires, which are considered as defeasible
belief and desire rules, a priority ordering on the desire rules,
as well as a set of facts and an initial decision (or prior
intentions). We assume that agents are autonomous, in the sense
that there are no priorities between desires of distinct agents.

\begin{definition}[Agent system specification]\label{def:ds}
An agent system specification is a tuple $AS=\langle
S,F,B,D,\geq,\delta^0\rangle$ that contains a set of agents $S$,
and for each agent $i$ a finite set of facts $F_{i} \subseteq
L_{W}\ (F=\bigcup_{i=1}^{n}F_i)$, a finite set of belief rules
$B_{i} \subseteq L_{AW}\times L_{W} \ (B=\bigcup_{i=1}^{n}B_i)$, a
finite set of desire rules $D_{i} \subseteq L_{AW}\times L_{AW} \
(D=\bigcup_{i=1}^{n}D_i)$, a relation $\geq_i \subseteq D_{i}
\times D_{i} \ (\geq=\bigcup_{i=1}^{n}\geq_i)$ which is a total
ordering (i.e. reflexive, transitive, and antisymmetric and for
any two elements $d_1$ and $d_2$ in $D_i$, either $d_1 \geq_i d_2$
or $d_2 \geq_i d_1$), and a finite initial decision $\delta^0_{i}
\subseteq L_A \ (\delta^0=\bigcup_{i=1}^{n}\delta^0_{i})$. For an
agent $i\in S$ we write $x\implies_i y$ for one of its rules.
\end{definition}

A belief rule `the agent $\alpha_i$ believes $y$ in context $x$'
is an ordered pair $x\implies_i y$ with $x\in L_{AW}$ and $y\in
L_W$, and a desire rule `the agent desires $y$ in context $x$' is
an ordered pair $x\implies_i y$ with $x\in L_{AW}$ and $y\in
L_{AW}$. It implies that the agent's beliefs are about the world
($x\implies_i p$), and not about the agent's decisions. These
beliefs can be about the effects of decisions made by the agent
($a\implies_i p$) as well as beliefs about the effects of
parameters set by the world ($p\implies_i q$). Moreover, the
agent's desires can be about the world ($x\implies_i p$,
desire-to-be), but also about the agent's decisions ($x \implies_i
a$, desire-to-do). These desires can be triggered by parameters
set by the world ($p\implies_i y$) as well as by decisions made by
the agent ($a\implies_i y$). Modelling mental attitudes such as
beliefs and desires in terms of defeasible rules results in what
might be called conditional mental attitudes \cite{broersen:aa01}.

\subsection{Agent Decisions}

The belief rules are used to determine the expected consequences
of a decision, where a decision $\delta$ is any subset of $L_A$
that contains the initial decision $\delta^0$. The set of expected
consequences of this decision $\delta$ is the belief extension of
$F\cup \delta$. Moreover, we consider a feasible decision as a
decision that does not imply a contradiction.

\begin{definition}[Decisions] \label{def:d}
Let $AS=\langle \{\alpha_1,\ldots,\alpha_n\},F,B,D,\geq,\delta^0
\rangle$ be an agent system specification. An $AS$ decision
profile $\delta$ for agents $\alpha_1$, \ldots, $\alpha_n$ is
$\delta=\langle \delta_1,\ldots \delta_n\rangle$ where $\delta_i$
is a decision of agent $\alpha_i$ such that
$$\delta^0_i \subseteq \delta_i \subseteq L_{A_i}\
\mbox{ for } i=1\ldots n$$ A feasible decision for agent $i$ is
$\delta_i$ such that
$$E_{B_i}(F_i\cup \delta_i) \mbox{ is consistent }$$
A feasible decision profile is a decision profile such that
$$E_B(F\cup \delta) \mbox{ is consistent }$$
where we write $E_B(F\cup \delta)$ for $\bigcup_{i=1}^{n}
E_{B_i}(F_i\cup \delta_i)$.
\end{definition}

The following example illustrates the decisions of a single agent.

\begin{example}
\label{ex:intro} Let $A_1=\{a,b,c,d,e\}$, $W=\{p,q\}$ and
$AS=\langle \{\alpha_1\},F,B,D,\geq,\delta^0\rangle$ with
$F_1=\{\neg p\}$, $B_1=\{c\implies q,d\implies q,e\implies \neg
q\}$, $D_1=\{\top\implies a,\top\implies b,b\implies
p,\top\implies q,d\implies q\}$, $\geq_1=\{b\implies
p>\top\implies b\}$, and $\delta^0_1=\{a\}$. The initial decision
$\delta^0_1$ reflects that the agent has already decided in an
earlier stage to reach the desire $\top\implies a$. Note that the
consequents of all $B_1$ rules are sentences of $L_W$, whereas the
antecedents of the $B_1$ rules as well as the antecedents and
consequents of the $D_1$ rules are sentences of $L_{AW}$.
We have due to the definition of $E_R(S)$:\\
$E_B(F\cup \{a\})=\{\neg p,a\}$ \\
$E_B(F\cup \{a,b\})=\{\neg p,a, b\}$ \\
$E_B(F\cup \{a,c\})=\{\neg p,a, c,q\}$ \\
$E_B(F\cup \{a,d\})=\{\neg p,a,d,q\}$ \\
$E_B(F\cup \{a,e\})=\{\neg p,a,e,\neg q\}$ \\
\ldots \\
$E_B(F\cup \{a,d,e\})=\{\neg p,a,d,e,q,\neg q\}$ \\
\ldots \\
Therefore $\{a,d,e\}$ is not a feasible $AS$ decision profile,
because its belief extension is inconsistent. Continued in Example
\ref{ex:intro2}.
\end{example}

The following example illustrates that the set of feasible
decisions of an agent may depend on the decisions of other agents.

\begin{example}
Let $A_1=\{a\}$, $A_2=\{b\}$, $W=\{p\}$ and $AS=\langle
\{\alpha_1,\alpha_2\},F,B,D,\geq,\delta^0\rangle$ with
$F_1=F_2=\emptyset$, $B_1=\{a\implies p\}$, $B_2=\{b\implies \neg
p\}$, $D_1=\{\top\implies p\}$, $D_2=\{\top\implies \neg p\}$,
$\geq$ is the identity relation, and
$\delta^0_1=\delta^0_2=\emptyset$. We have that $\langle
\emptyset,\{b\}\rangle$ is a feasible decision profile, but
$\langle \{a\},\{b\}\rangle$ is not. If $\delta_1=\emptyset$, then
agent $\alpha_2$ can decide $\delta_2=\{b\}$. However, if
$\delta_1=\{a\}$, then agent $\alpha_2$ cannot decide so, i.e.
$\delta_2\not=\{b\}$.
\end{example}

\subsection{Agent preferences}
\label{sec:ndt}

In this section we introduce a way to compare decisions. We
compare decisions by comparing sets of desire rules that are not
reached by the decisions. Since only ordering on individual desire
rules, and not ordering on sets of desire rules, are given, we
first lift the ordering on individual desire rules to an ordering
on sets of desire rules.

\begin{definition} \label{def:geq}
Let $AS=\langle S,F,B,D,\geq,\delta^0\rangle$ be an agent system
specification, $D'_i,D''_i$ two subsets of $D_i$, and $D_i
\setminus D'_i$ the set of $D_i$ elements which are not $D'_i$
elements. We have $D'_i\succeq D''_i$ if $\forall d'' \in
D''_i\setminus D'_i \exists d' \in D'_i \setminus D''_i \mbox{
such that } d' > d''$. We write $D'_i\succ D''_i$ if $D'_i\succeq
D''_i$ and $D''_i \not \succeq D'_i$, and we write $D'_i\simeq
D''_i$ if $D'_i\succeq D''_i$ and $D''_i\succeq D'_i$.
\end{definition}

The following propositions show that the priority relation
$\succeq$ is reflexive, anti-symmetric, and transitive.

\begin{proposition}
Let $D_1 \not \subseteq D_3$ and $D_3 \not \subseteq D_1$. For
finite sets, the relation $\succeq$ is reflexive ( $\forall D \ D
\succeq D$), anti-symmetric $\forall D_1 , D_2 \ (D_1 \succeq D_2
\ \wedge D_2 \succeq D_1) \rightarrow D_1 = D_2$, and transitive,
i.e. $D_1 \succeq D_2$ and $D_2 \succeq D_3$ implies $D_1 \succeq
D_3$.
\end{proposition}

The desire rules are used to compare the decisions. The comparison
is based on the set of unreached desires and not on the set of
violated or reached desires. A desire $x\implies y$ is unreached
by a decision if the expected consequences of this decision imply
$x$ but not $y$. The desire rule is violated or reached if these
consequences imply respectively $x\wedge \neg y$ or $x\wedge y$,
respectively.

\begin{definition}[Comparing decisions] \label{def:cd}
Let $AS=\langle S,F,B,D,\geq,\delta^0\rangle$ be an agent system
specification and $\delta$ be a $AS$ decision. The unreached
desires of decision $\delta$ for agent $\alpha_i$
are:\\
\\
$U_i(\delta)=\\ \{x\implies y \in  D_i \mid E_{B}(F\cup \delta)
\models x \mbox{ and } E_{B}(F\cup \delta) \not
\models y\}$\\
\\
Decision $\delta$ is at least as good as decision $\delta'$ for
agent $\alpha_i$, written as $\delta\geq^U_i \delta'$, iff
$$U_i(\delta') \succeq U_i(\delta)$$
Decision $\delta$ dominates decision $\delta'$ for agent
$\alpha_i$, written as $\delta>^U_i \delta'$, iff
$$\delta\geq^U_i \delta'\mbox{ and } \delta' \not \geq^U_i \delta$$
\end{definition}

The following continuation of Example \ref{ex:intro} illustrates
the comparison of decisions.

\begin{example}[Continued]\label{ex:intro2}
We have:\\
$U(\{a\})=\{\top \implies b,\top\implies q\}$, \\
$U(\{a,b\})=\{b \implies p,\top\implies q\}$, \\
$U(\{a,c\})=\{\top \implies q\}$,\\
$U(\{a,d\})=\{\top \implies b,d\implies q\}$,\\
$U(\{a,e\})=\{\top \implies b,\top\implies q\}$,\\
$U(\{a,b,c\})=\{b \implies p\}$.\\
\ldots\\
We thus have for example that the decision $\{a,c\}$ dominates the
initial decision $\{a\}$, i.e. $\{a,c\}>^U\{a\}$. There are two
decisions for which their set of unreached contains only one
desire. Due to the priority relation, we have that
$\{a,c\}>^U\{a,b,c\}$.
\end{example}

\subsection{Agent games}

In this subsection, we consider agents interactions based on agent
system specifications, their corresponding agent decisions, and
the ordering on the decisions as explained in previous
subsections. Game theory is the usual tool to model the
interaction between self-interested agents. Agents select optimal
decisions under the assumption that other agents do likewise. This
makes the definition of an optimal decision circular, and game
theory therefore restricts its attention to equilibria. For
example, a decision is a Nash equilibrium if no agent can reach a
better (local) decision by changing its own decision. The most
used concepts from game theory are Pareto efficient decisions,
dominant decisions and Nash decisions. We first repeat some
standard notations from game theory
\cite{binmore92fun,osborne94course}.

As mentioned, we use $\delta_i$ to denote a decision of agent
$\alpha_i$ and $\delta=\langle \delta_1,\ldots,\delta_n \rangle$
to denote a decision profile containing one decision for each
agent. $\delta_{-i}$ is the decision profile of all agents except
the decision of agent $\alpha_i$. $( \delta_{-i} ,\delta'_{i} )$
denotes a decision profile which is the same as $\delta$ except
that the decision of agent $i$ from $\delta$ is replaced with the
decision of agent $i$ from $\delta'$. $\delta'_i >^U_{i} \delta_i$
denotes that decision $\delta'_i$ is better than $\delta_i$
according to his preferences $>^U_{i}$ and $\delta'_i \geq^U_{i}
\delta_i$ if better or equal. $\Delta$ is the set of all decision
profiles for agents $\alpha_1, \ldots,\alpha_n$, $\Delta_f
\subseteq \Delta$ is the set of feasible decision profiles, and
$\Delta^i$ is the set of possible decisions for agent $\alpha_i$.

\begin{definition}[Game specification]
Let $AS=\langle
S=\{\alpha_1,\ldots,\alpha_n\},F,B,D,\geq,\delta^0\rangle$ be
specification of agent system in $S$, $A_i$ be the set of $AS$
feasible decisions of agent $\alpha_i$ according to Definition
\ref{def:d}, $\Delta_{f}=A_1 \times \ldots \times A_n$, and
$\geq^U_i$ be the $AS$ preference relation of agent $\alpha_i$
defined on its feasible decisions according to definition
\ref{def:cd}. Then, the game specification of $AS$ is the tuple
$\langle S,\Delta_{f},(\geq^U_i)\rangle$.
\end{definition}

We now consider different types of decision profiles which are
similar to types of strategy profiles from game theory.

\begin{definition}
A PS decision profile $\delta=\langle \delta_1, \ldots, \delta_n
\rangle \in \Delta_f$ is:
\begin{description}
\item[Pareto decision]
if there is no $\delta'=\langle\delta'_1, \ldots, \delta'_n\rangle
\in \Delta_f$ for which $\delta'_i >^U_{i} \delta_i$ for all
agents $\alpha_i$.

\item[strongly Pareto decision]
if there is no $\delta'=\langle\delta'_1, \ldots, \delta'_n\rangle
\in \Delta_f$ for which $\delta'_i \geq^U_{i} \delta_i$ for all
agents $\alpha_i$ and $\delta'_j >^U \delta_j$ for some agents
$\alpha_j$.

\item[dominant decision]
if for all $\delta' \in \Delta_f$ and for every agent $i$ it
holds: $( \delta'_{-i} , \delta_{i} ) \geq^U_{i} (\delta'_{-i} ,
\delta'_{i} )$ i.e. a decision is dominant if it yields a better
payoff than any other decisions regardless of what the other
agents decide.

\item[Nash decision]
if for all agents i it holds: $(\delta_{-i} , \delta_{i})
\geq^U_{i} (\delta_{-i} , \delta'_{i}) \mbox{ for all }
\delta'_{i} \in \Delta_f^{i}$
\end{description}
\end{definition}

It is a well known fact that Pareto decisions exist (for finite
games), whereas dominant decisions do not have to exist. The
latter is illustrated by the following example.

\begin{example}
Let $\alpha_1$ and $\alpha_2$ be two agents, $F_1=F_2=\emptyset$,
and initial decisions $\delta^0_1=\delta^0_2=\emptyset$. They have the following beliefs en desires:\\
$B_{\alpha_1}=\{a \Rightarrow p, \neg a \Rightarrow \neg p \}$ \\
$D_{\alpha_1}=\{\top \Rightarrow p, \top \Rightarrow q \}$ \\
$\geq_{\alpha_1}= \top \Rightarrow p > \top \Rightarrow q > \top \Rightarrow \neg q > \top \Rightarrow \neg p$ \\
$B_{\alpha_2}=\{b \Rightarrow q, \neg b \Rightarrow \neg q \}$ \\
$D_{\alpha_2}=\{\top\Rightarrow \neg p, \top\Rightarrow \neg q \}$ \\
$\geq_{\alpha_2}= \top \Rightarrow \neg p > \top \Rightarrow \neg q > \top \Rightarrow q > \top \Rightarrow p$ \\
\\
Let $\Delta_f$ be feasible decision profiles, $E_B$ be the
outcomes of the decisions, and $U(\delta_{i})$ be the set of
unreached desires for agent $\alpha_i$. \\
\\
$\begin{array}{lllll}
\Delta & E_B & U_{\delta_{1}} & U_{\delta_{2}}\\
\\
\langle a,b \rangle & \{p,q\} & \emptyset & \{\top \Rightarrow
\neg
p, \\
&&&\ \ \top \Rightarrow \neg q \}\\
\langle a,\neg b \rangle & \{p,-q\} & \{\top \Rightarrow q\} &
\{\top
\Rightarrow \neg p \}\\
\langle \neg a,b\rangle & \{\neg p,q \} & \{\top \Rightarrow p\} &
\{\top \Rightarrow \neg q \}\\
\langle \neg a, \neg b \rangle & \{\neg p,\neg q \} & \{\top
\Rightarrow p , \top \Rightarrow q \} & \emptyset
\end{array}$\\
\\
\\
According to definition 5, for $A_1 :\\U(\langle a,b \rangle) >
U(\langle a, \neg b \rangle)>U(\langle \neg a,b \rangle)>U(\langle
\neg a,\neg b \rangle)$\\ and for $A_2: \\U(\langle \neg a,\neg b
\rangle)>U(\langle \neg a,b \rangle)>U(\langle a,\neg b
\rangle)>U(\langle a,b \rangle)$.\\ None of these decision
profiles are dominant decisions, i.e. the agents specifications
has no dominant solution with respect to their unreached desires.
\end{example}

The following example illustrates a typical cooperation game.
\begin{example}
$B_1=\{a\implies p \ , \ b \implies \neg p \wedge q\}$,
$D_1=\{\top \implies p\wedge q\}$ $B_2=\{c\implies q \ , \ d
\implies p \wedge \neg q\}$ $D_2=\{\top \implies p\wedge q\}$. The
agents have a common goal $p\wedge q$, which they can only reach
by cooperation.
\end{example}

The following example illustrates a qualitative version of the
notorious prisoner's dilemma, where the selfish behavior of
individual autonomous agents leads to global bad decisions.

\begin{example}
Let $A_1=\{a\}$ ($\alpha_1$ cooperates), $A_2=\{b\}$ ($\alpha_2$
cooperates), and $AS$ be an agent system specification with
$D_1=\{\top \implies \neg a\wedge b, \top \implies b,\top\implies
\neg (a\wedge \neg b)\}$, $D_2=\{\top \implies a\wedge \neg b,
\top \implies a,\top\implies \neg (\neg a\wedge b)\}$. The only
Nash decision is $\{\neg a,\neg b\}$, whereas both agents would
prefer $\{a,b\}$.
\end{example}

Starting from an agent system specification, we can derive the
game specification and in this game specification we can use
standard techniques to for example find the Pareto decisions.
However, the problem with this approach is that the translation
from an agent system specification to a game specification is
computationally expensive. For example, a compact agent
representation with only a few belief and desire rules may lead to
a huge set of decisions if the number of decision variables is
high.

The main challenge of qualitative game theory is therefore whether
we can bypass the translation to game specification, and define
properties directly on the agent system specification. For
example, are there particular properties of agent system
specification for which we can prove that there always exists a
dominant decision for its corresponding derived game
specification? A simple example is an agent system specification
in which each agent has the same belief and desire rules.

In this paper we do not further pursue these issues, but we turn
to our focus of interest: joint goals.

\section{Joint goals} \label{sec:gdt}

In this section we ask ourselves the question whether and how we
can interpret a decision profile or equilibrium as goal-based or
goal-oriented behavior.
We first define decision rules and sets of decision profiles
closed under indistinguishable decision profiles.

\subsection{Decision rule}
A decision rule maps a agent system specification to a set of
possible decision profiles.

\begin{definition}
A decision rule is a function from agent system specifications to
sets of feasible decision profiles.
\end{definition}

Decision theory prescribes a decision maker to select the optimal
or best decision, which can be defined as a decision that is not
dominated. Is there an analogous prescription for societies of
agents? A set of {\em cooperating} agents has to select an optimal
or Pareto decision. We call such cooperating agents a BD rational
society of cooperating agents.
\begin{definition}\label{def:ra}
A BD rational society of cooperating agents is a set of agents,
defined by an agent system specification $AS$, that selects a
Pareto $AS$ decision.
\end{definition}

In this paper we consider decision rules based on unreached
desires. We therefore assume that a decision rule cannot
distinguish between decision profiles $\delta_1$ and $\delta_2$
such that $\delta_1\sim^U \delta_2$. We say that two decision
profiles $\delta$ and $\delta'$ are indistinguishable if
$U(\delta)=U(\delta')$, and we call a set of decisions U-closed if
the set is closed under indistinguishable decision profiles.
\begin{definition}
Let $AS=\langle S,F,B,D,\geq,\delta^0\rangle$ be an agent system
specification and $\Delta$ a set of $AS$ decision profiles.
$\Delta$ is U-closed if $\delta\in \Delta$ implies $\delta'\in
\Delta$ for all $AS$ decision profiles $\delta'$ such that
$U(\delta)=U(\delta')$.
\end{definition}

An example of a decision rule is the function that maps agent
system specifications to Pareto decision profiles (a BD rational
decision rule). Another example is a function that maps agent
system specifications to Nash equilibria if they exist, otherwise
to Pareto decisions.

\subsection{Goals}

In this section we show that every society of agents can be
understood as planning for joint goals, whether the decision is
reached by cooperation or is given by a (e.g. Nash) equilibrium.
We define not only goals which must be reached, called positive
goals, but we add negative goals. Negative goals are defined in
the following definition as states the agent has to avoid. They
function as constraints on the search process of goal-based
decisions.

\begin{definition}[Goal-based decision] \label{def:gd2}
Let $AS=\langle S,F,B,D,\geq,\delta^0\rangle$ be an agent system
specification, and the so-called positive joint goal set $G^+$ and
the negative joint goal set $G^-$ be subsets of $L_{AW}$. A
decision $\delta$ is a $\langle G^+,G^-\rangle$ decision if
$E_B(F\cup \delta)\models_{AW} G^+$ and for each $g\in G^-$ we
have $E_B(F\cup \delta)\not\models_{AW} g$.
\end{definition}

Joint goals are defined with respect to a set of decision
profiles. The definition of $\Delta$ joint goal set encodes a
decision profile together with its indistinguishable decision
profiles as a positive and negative goal set.

\begin{definition}[$\Delta$ goal set]
Let $AS=\langle S,F,B,D,\geq,\delta^0\rangle$ be an agent system
specification and $\Delta$ an U-closed set of feasible decisions.
The two sets of formulas $\langle G^+,G^-\rangle\subseteq
(L_{AW},L_{AW})$ is a $\Delta$ joint goal set of $AS$ if there is
an $AS$ decision $\delta\in \Delta$ such that
$$G^+=\{y\mid x\implies y\in D, E_{B}(F \cup \delta)\models_{AW} x\wedge y\}$$
$$G^-=\{x\mid x\implies y\in D, E_{B}(F\cup d)\not\models_{AW}x\}$$
$\langle G^+,G^-\rangle\subseteq (L_{AW},L_{AW})$ is a feasible
joint goal set of $AS$ if there is an U-closed set of feasible
decisions $\Delta$ such that $\langle G^+,G^-\rangle\subseteq
(L_{AW},L_{AW})$ is a $\Delta$ joint goal set of $AS$. $\langle
G^+,G^-\rangle\subseteq (L_{AW},L_{AW})$ is a joint goal set of
$AS$ if there is a set $D'\subseteq D$ such that
$$G^+=\{y\mid x\implies y\in D'\}$$
$$G^-=\{x\mid x\implies y\in D'\}$$
\end{definition}

The first part of the representation theorem follows directly from
the definitions.

\begin{proposition}\label{prop:mg1}
Let $AS=\langle S,F,B,D,\geq,\delta^0\rangle$ be an agent system
specification and $\Delta$ be an U-closed set of feasible decision
profiles. For a decision profile $\delta\in \Delta$ of $AS$ there
is a $\Delta$ joint goal set $\langle G^+,G^-\rangle$ of $AS$ such
that $\delta$ is a $\langle G^+,G^-\rangle$ decision.

\proof\ Follows directly from the definitions.
\end{proposition}

\begin{proposition}\label{prop:mg2}
Let $AS=\langle S,F,B,D,\geq,\delta^0\rangle$ be an agent system
specification and $\Delta$ an U-closed set of feasible decision
profiles. For a $\Delta$ joint goal set $\langle G^+,G^-\rangle$
of $AS$, a $\langle G^+,G^-\rangle$ decision is a $\Delta$
decision.

\proof\ Follows from U-closed property.
\end{proposition}

The representation theorem is a combination of Proposition
\ref{prop:mg1} and \ref{prop:mg2}.

\begin{theorem}
Let $AS=\langle S,F,B,D,\geq,\delta^0\rangle$ be an agent system
specification and $\Delta$ an U-closed set of feasible decision
profiles. A decision profile $\delta$ is in $\Delta$ if and only
if there is a $\Delta$ goal set $\langle G^+,G^-\rangle$ of $AS$
such that $\delta$ is a $\langle G^+,G^-\rangle$ decision profile.
\end{theorem}

The second theorem follows from the first one.
\begin{theorem}
Let $AS=\langle S,F,B,D,\geq,\delta^0\rangle$ be an agent system
specification. A decision profile $\delta$ is a feasible $AS$
decision profile if and only if there is a feasible goal set
$\langle G^+,G^-\rangle$ of $AS$ such that $\delta$ is a $\langle
G^+,G^-\rangle$ decision profile.
\end{theorem}

Consider a society of agent that tries to determine its Pareto
decision profiles. The game specifications suggest the following
algorithm:
\begin{verbatim}
 calculate all decision profiles
 for all decision profiles,
   calculate consequences
 order all decision profiles
\end{verbatim}

The goal-based representation suggest an alternative approach:
\begin{verbatim}
 calculate all joint goals
 filter feasible joint goals
 for each feasible joint goal set
   find goal-based decision profiles
 order these decision profiles
\end{verbatim}

In other words, the goal-based representation suggests to
calculate the joint goals first. However, the problem to calculate
these joint goals is still computationally hard. There are two
ways to proceed:
\begin{itemize}
\item
Define heuristics for the optimization problem;
\item
Find a fragment of the logic, such that the optimization becomes
easier.
\end{itemize}

For example, consider the following procedure to find (positive)
goals:
$$G\subseteq E_{B\cup D}(F\cup \delta^0)$$
This procedure is not complete, because it does not take effects
of actions into account. Thus, in the general case it can be used
as a heuristic. Moreover, it is complete for the fragment in which
the belief rules do not contain effects of actions, i.e.
$B\subseteq S\times L_W\times L_W$.\footnote{We cannot add beliefs
on decision variables, as will be clear in the next section.
Suppose $A=\{a\}$ and $B=\{\top \rightarrow a\}$. Clearly all
decisions should be considered. However, all decision profiles
given in Definition 3 imply a and thus all decisions not implying
a would be excluded.}

\section{Concluding remarks}

In this paper we have defined a qualitative decision and game
theory in the spirit of classical decision and game theory. The
theory illustrates the micro-macro dichotomy by distinguishing the
optimization problem from game theoretic equilibria. We also
showed that any group decision, whether based on optimization or
on an equilibrium, can be represented by positive and negative
goals.

We think that the method of this paper is more interesting than
its formal results. The decision and game theory are based on
several ad hoc choices which need further investigation. For
example, the desire rules are defeasible but the belief rules are
not (the obvious extension leads to wishful thinking problems as
studied in \cite{thomason:kr00,broersen:ecsqaru01}). However, the
results suggest that any group decision can be understood as
reaching for goals. We hope that further investigations along this
line brings the theories and tools used for individual agents and
multi agent systems closer together.

\bibliographystyle{plain}
\bibliography{goalplan}

\end{document}